\documentclass[letterpaper,english,preprint, aps, prl, showpacs]{revtex4}
\usepackage[latin1]{inputenc}
\usepackage{graphicx}

\makeatletter
\usepackage{babel}
\makeatother
\begin{document}

\title{Multiprotein DNA looping}

\author{Jose M. G. Vilar}

\email{vilar@cbio.mskcc.org}

\author{Leonor Saiz}

\email{leonor@cbio.mskcc.org}

\affiliation{Integrative Biological Modeling Laboratory, Computational Biology
Program, Memorial Sloan-Kettering Cancer Center, New York, NY 10021}

\begin{abstract}
DNA looping plays a fundamental role in a wide variety of biological
processes, providing the backbone for long range interactions on DNA.
Here we develop the first model for DNA looping by an arbitrarily
large number of proteins and solve it analytically in the case of
identical binding. We uncover a switch-like transition between looped
and unlooped phases and identify the key parameters that control this
transition. Our results establish the basis for the quantitative understanding
of fundamental cellular processes like DNA recombination, gene silencing,
and telomere maintenance.
\end{abstract}


\maketitle
The formation of DNA loops by the binding of proteins and protein
complexes at distal DNA sites plays a fundamental role in many cellular
processes \cite{reviews adhya,reviews schleif,ours,matthews,DNABP},
including transcription \cite{roeder}, recombination \cite{broach},
replication \cite{matthews}, and telomere maintenance \cite{de Lange}.
Disruption or alteration of these processes often results in different
developmental disorders and disease states, with cancer the most prominent
example \cite{weinberg}. The key role of looping is to bypass the
one dimensional nature of DNA and allow distal DNA sites to come close
to each other. In gene regulation, proteins bound far away from the
genes they regulate can be brought to the initiation of transcription
region of the regulated genes by looping the intervening DNA \cite{ours}.
Similarly, in DNA recombination, loops are formed that bring together
two DNA regions to transfer the genetic information from one DNA region
to another. Although there are studies of double-stranded DNA looping
by DNA itself (cyclization) \cite{marko}, by one protein \cite{vilar&leibler,schulten},
or by a few proteins \cite{ours}, a general understanding of the
collective properties that might emerge when multiple proteins are
involved is still lacking. The case of multiple proteins is specially
important because it is the dominant one for loops larger than a few
hundred base pairs \cite{ours,matthews}. 

In this letter we develop the first model for DNA looping by an arbitrary
number of proteins. For small number of proteins, this model accounts
for previous thermodynamic approaches that have been shown to reproduce
in detail available experimental data on regulation of the \emph{lac}
operon and phage-$\lambda$ \cite{ours}. For large number of proteins,
we show here that the model exhibits properties reminiscent of phase
transitions \cite{stanley}, with a quasi-discontinuity in the occupancy
of the DNA sites by DNA binding proteins. We identify the parameters
that control the transition and show that there are two phases that
can be associated with looped and unlooped states of DNA. The density
of proteins on DNA is low for the unlooped state and high for the
looped state. Despite the apparent one dimensional physical nature
of the problem, looping of DNA introduces long range interactions
which make the system exhibit unexpected collective features. 

\begin{figure}[p]
\includegraphics[%
  scale=0.8]{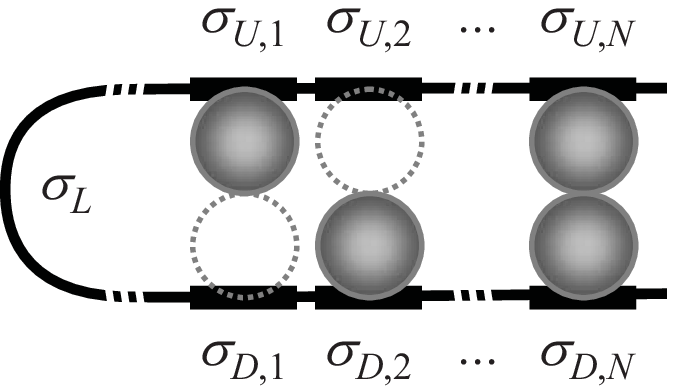}

\caption{Schematic representation of looped DNA. Proteins (filled circles)
bind to DNA (black line) at specific sites (rectangles on the line).
Proteins bound at one operator, upstream (\emph{U}) or downstream
(\emph{D}), can interact with their counterparts at the opposite operator
if DNA forms a loop (\emph{L}). In this example, the number of binding
sites per operator is $N$. The binary variables $\sigma_{U,i}$ and
$\sigma_{D,i}$ are 1 when proteins are bound to the corresponding
DNA site and are 0 otherwise. Here, only the two proteins bound at
sites $i=N$ on the upstream (\emph{U},$N$) and downstream (\emph{D},$N$)
operator interact with each other.\label{cap:System}}
\end{figure}

We consider a system with two spatially distinct DNA regions on the
same DNA double-strand, referred to as upstream (\emph{U}) and downstream
(\emph{D}) operators (Figure~\ref{cap:System}). Each operator has
$N$ binding sites for proteins that once bound to one of them can
interact with its symmetric counterpart on the other operator if DNA
is looped. The typical way to obtain the statistical properties of
the system is to identify the representative states and their corresponding
free energies and to compute the partition function \cite{vilar&leibler}.
This process is usually done by tabulating the free energies and explicitly
writing down the sums of Boltzmann factors for all the states. For
large systems, however, this procedure is not practical because of
the exponential growth of the potential number of states (e.g., for
N=3, there are already 128 states).

The facts that the free energy of a state can be decomposed into different
contributions \cite{ours} and that the states can be labelled by
discrete variables \cite{complejos} allow for a Hamiltonian description
of the system. Here, we describe the binding of proteins to DNA through
binary variables $\sigma_{U,i}$ and $\sigma_{D,i}$, which indicate
whether ($=1$) or not ($=0$) a protein is bound to site $i$ at
the upstream or downstream operator, respectively. Similarly, an additional
binary variable $\sigma_{L}$ indicates whether DNA is looped ($=1$)
or not ($=0$). In terms of this set of binary variables the system
is described by the following Hamiltonian:\begin{equation}
H=(c+e\sum_{i=1}^{N}\sigma_{U,i}\sigma_{D,i})\sigma_{L}+g\sum_{i=1}^{N}(\sigma_{U,i}+\sigma_{D,i})\;,\label{eq:Hamiltonian}\end{equation}
 where $g$ is the change in free energy upon binding of a protein
to a DNA site; $e$ is the free energy of interaction between proteins
symmetrically bound at opposite operators; and $c$ is the free energy
of forming the DNA loop \cite{ours,pnas}. Therefore, the free energy
of each of the $2^{2N}$ looped and $2^{2N}$ unlooped states is obtained
directly from the previous Hamiltonian. The dependence of the Hamiltonian
on the concentration of binding proteins $n$ enters, in the usual
form, through the quantity $g$, which can be viewed as a chemical
potential: $g=g^{o}-\frac{1}{\beta}\ln n$, where $g^{o}$ denotes
the value of $g$ at a protein concentration of 1 M and $\beta^{-1}=RT$
(the gas constant times the absolute temperature). This type of Hamiltonians
account for thermodynamic models that have recently been shown to
accurately describe gene regulation in the \emph{lac} operon by the
\emph{lac} repressor ($N=1$) and in phage-$\lambda$ by the cI$_{2}$
repressor ($N=3$) \cite{complejos,ours}. A systematic analysis for
large systems, however, is still missing. 

\begin{figure}
\includegraphics[%
  scale=0.50]{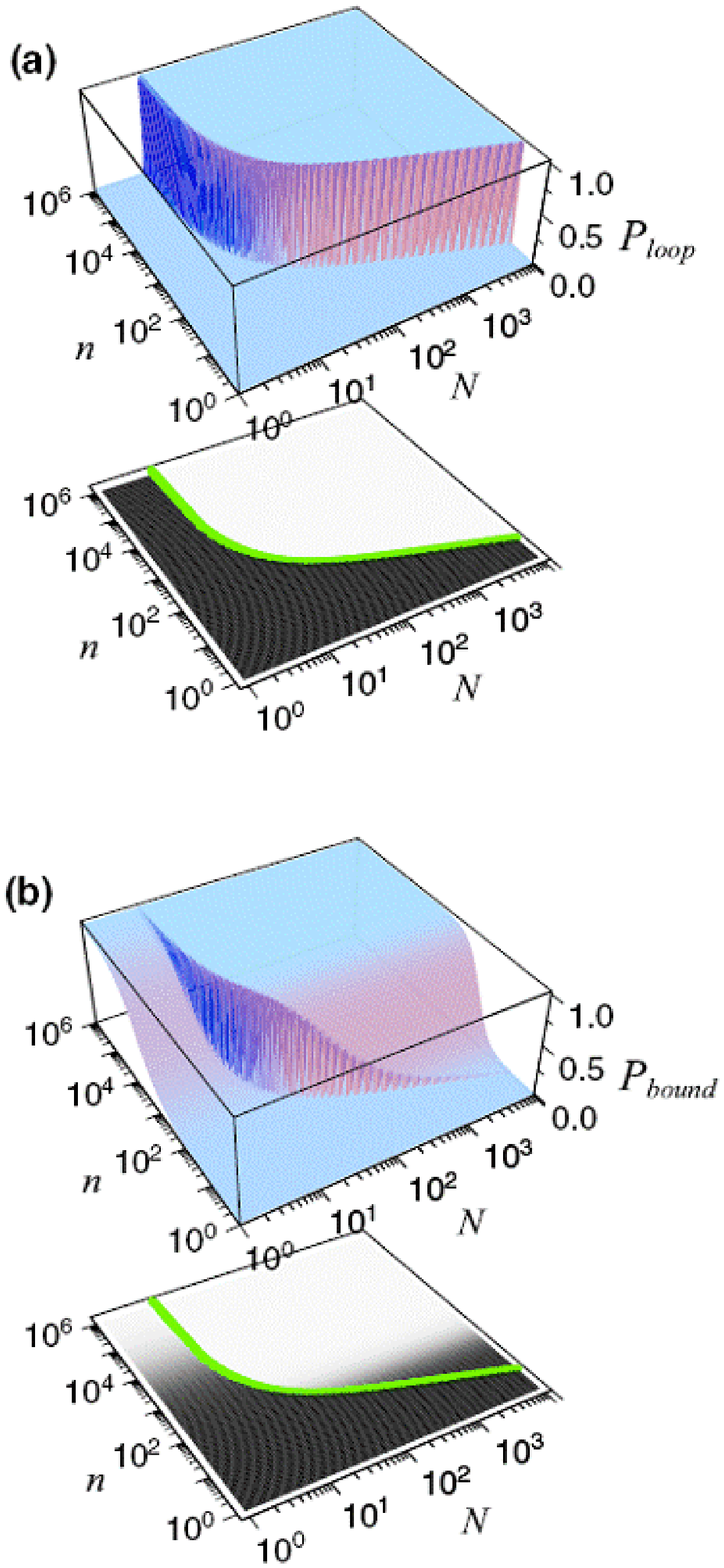}

\caption{Looping probability $P_{loop}$ (a) and site occupancy $P_{bound}$
(b) as functions of the protein concentration $n$ (in nM) and the
number of binding sites per operator $N$. The values of the parameters
are $\beta^{-1}=0.6$ kcal/mol, $g^{o}=-7.2$ kcal/mol, $c=30$ kcal/mol,
$e=-5.5$ kcal/mol. Black and white colors in the 2-D density plot
projections of the 3-D surfaces represent probabilities 0 and 1, respectively.
The green line corresponds to $\tilde{n}$ (given by Equation \ref{eq:Sol-n-general})
and indicates the separation between looped and unlooped phases (regions
with concentrations above and below the line, respectively). Note
that there is no looped phase for $N<-c/e=5.45$.\label{cap:3D}}
\end{figure}

In order to compute the partition function, it is convenient to rewrite
the Hamiltonian as the sum of quasi-independent single-pair Hamiltonians:\begin{equation}
H=\sum_{i=1}^{N}H_{P,i}\;,\label{eq:sumsinglepair}\end{equation}
 where \begin{equation}
H_{P,i}=\sigma_{L}(c/N+e\sigma_{U,i}\sigma_{D,i})+g(\sigma_{U,i}+\sigma_{D,i})\;.\label{eq:singlepair}\end{equation}
The coupling of single-pair Hamiltonians is established through the
three-body terms $e\sigma_{U,i}\sigma_{D,i}\sigma_{L}$, which account
for the interactions between DNA looping and DNA-bound proteins. 

The quasi-independence property allows us to express the partition
function as\begin{equation}
Z=\sum_{\sigma_{L}=\{0,1\}}\prod_{i=1}^{N}Z_{P,i}\;,\label{eq:partitionfunction}\end{equation}
 with \begin{eqnarray}
Z_{P,i} & = & \sum_{{{\sigma_{U,i}=\{0,1\}\atop \sigma_{D,i}=\{0,1\}}}}e^{-\beta H_{P,i}}\label{eq:partialZ}\\
 & = & e^{-\frac{c\beta\sigma_{L}}{N}}+e^{-\beta\left(2g+\left(\frac{c}{N}+e\right)\sigma_{L}\right)}+2e^{-\beta\left(g+\frac{c\sigma_{L}}{N}\right)}\;,\nonumber \end{eqnarray}
 which leads to\begin{eqnarray}
Z & = & \left(e^{-2g\beta}\left(1+e^{g\beta}\right)^{2}\right)^{N}\label{eq:explicitZ}\\
 &  & +\left(2e^{-\left(\frac{c}{N}+g\right)\beta}+e^{-\left(\frac{c}{N}+e+2g\right)\beta}+e^{-\frac{c\beta}{N}}\right)^{N}\;.\nonumber \end{eqnarray}

The two properties of interest are the looping probability and the
occupancy of the sites, which follow straightforwardly from the previous
expression of the partition function. The probability of the looped
state is given by the average value of $\sigma_{L}$, $\left\langle \sigma_{L}\right\rangle =-\frac{1}{\beta}\frac{\partial}{\partial c}\ln Z$.
After taking the logarithmic derivative and performing algebraic manipulations,
we obtain \begin{equation}
\left\langle \sigma_{L}\right\rangle =\frac{1}{1+X^{N}}\;,\label{eq:Loop-Hill-like}\end{equation}
 with\begin{equation}
X=\frac{e^{\left(\frac{c}{N}+e\right)\beta}\left(1+e^{g\beta}\right)^{2}}{1+2e^{(e+g)\beta}+e^{(e+2g)\beta}}\;.\label{eq:Loop-Hill-like-X}\end{equation}
 This expression for $\left\langle \sigma_{L}\right\rangle $ indicates
that, for large $N$, there is the potential for a sharp transition
between two states: the loop is always present if $X<1$ and absent
if $X>1$. This discontinuity can also propagate to the probability
for a site to be occupied, given by $\left\langle \sigma_{U/D,i}\right\rangle =-\frac{1}{2N\beta}\frac{\partial}{\partial g}\ln Z$
, which is related to the looping probability through\begin{eqnarray}
\left\langle \sigma_{U/D,i}\right\rangle  & = & \frac{1}{1+e^{g\beta}}\left\langle 1-\sigma_{L}\right\rangle \nonumber \\
 &  & +\frac{1+e^{(e+g)\beta}}{1+2e^{(e+g)\beta}+e^{(e+2g)\beta}}\left\langle \sigma_{L}\right\rangle \;.\label{eq:Occupancy}\end{eqnarray}

Under physiological conditions, the parameter typically used by the
cell to control DNA looping is the protein concentration. Figure \ref{cap:3D}
shows the system behavior as a function of the protein concentration
and the number of binding sites for representative values of the parameters
\cite{new ref}. The figure illustrates the presence of looped and
unlooped phases (Figure \ref{cap:3D}a). Only for intermediate concentrations
the occupancy of the sites (Figure \ref{cap:3D}b) displays a discontinuous
behavior. For concentrations in the high and low extremes, DNA looping
does not substantially affect the binding of proteins.

\begin{figure}
\includegraphics[%
  scale=0.7]{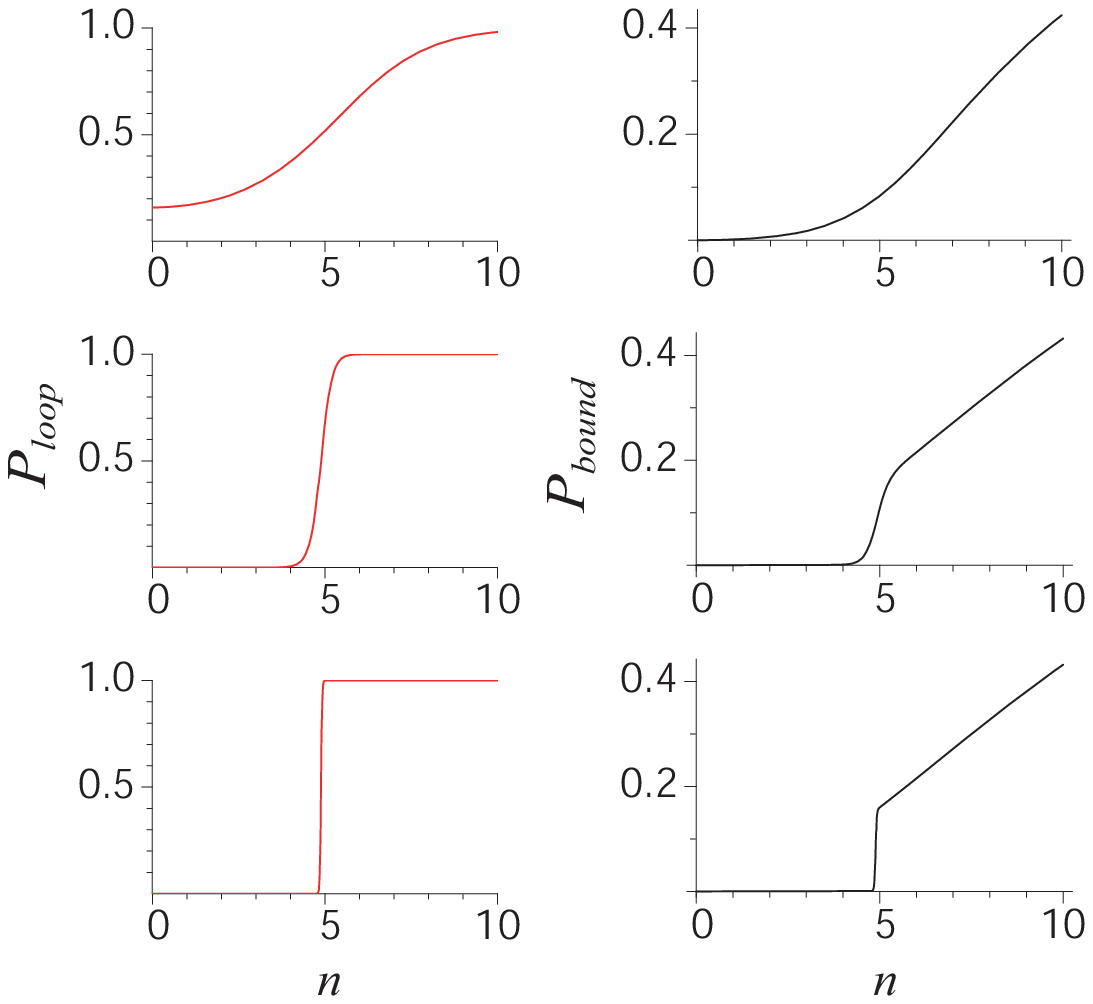}

\caption{Looping probability $P_{loop}$ and site occupancy $P_{bound}$ as
functions of the protein concentration $n$ (in nM) for coordinated
changes of the free energy of looping and number of binding sites.
The values of the parameters are $\beta^{-1}=0.6$ kcal/mol, $g^{o}=-7.2$
kcal/mol, $e=-7.5$ kcal/mol, $c=0.1N$ kcal/mol, $N=10$ (top), $N=100$
(middle), and $N=1000$ (bottom).\label{cap:TO}}
\end{figure}

\begin{figure}
\includegraphics[%
  scale=0.7]{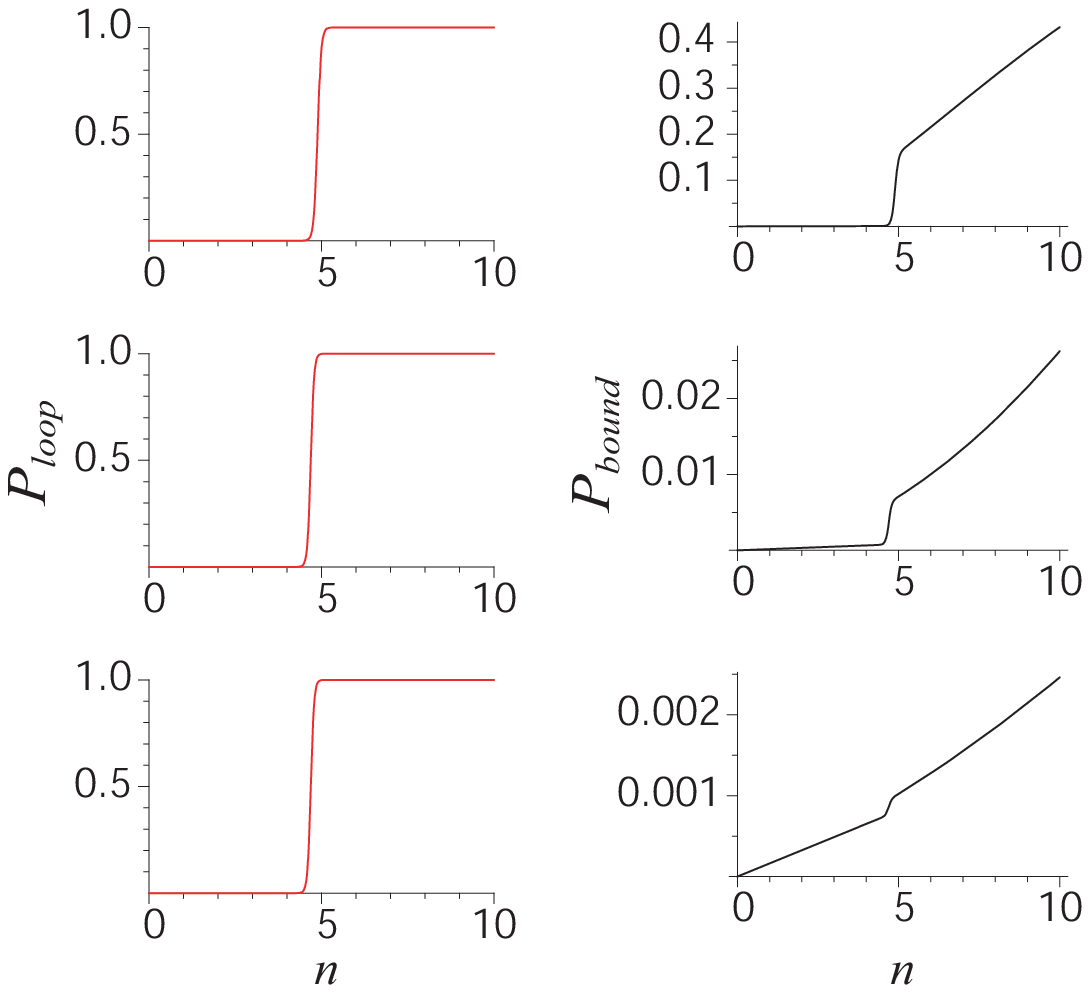}

\caption{Looping probability $P_{loop}$ and site occupancy $P_{bound}$ as
functions of the protein concentration $n$ (in nM) for coordinated
changes in the interaction free energy and number of binding sites.
The values of the parameters are $\beta^{-1}=0.6$ kcal/mol, $g^{o}=-7.2$
kcal/mol, $c=30$ kcal/mol, $e=-10.96\:\textrm{kcal/mol}+\beta^{-1}\ln N$,
$N=100$ (top), $N=1000$ (middle), and $N=10000$ (bottom).\label{cap:TS}}
\end{figure}

The concentration $\tilde{n}$ at which the transition happens ($X=1$)
is given by \begin{equation}
\left.\tilde{n}\right.=e^{\beta g^{o}}\frac{e^{e\beta}\left(e^{\frac{c\beta}{N}}-1\right)+\sqrt{e^{e\beta}\left(1-e^{e\beta}\right)\left(e^{\frac{c\beta}{N}}-1\right)}}{1-e^{\left(\frac{c}{N}+e\right)\beta}}\;.\label{eq:Sol-n-general}\end{equation}
 This equation has a positive solution if and only if $e<-c/N$. If
$e\ge-c/N$ there is no positive solution and the sites become occupied
as the concentration increases without the system ever reaching the
looped state (Figure \ref{cap:3D}). Therefore, the inter-operator
protein interactions need to exceed a strength threshold in order
for DNA looping to have the potential to be present. Remarkably, this
threshold goes to zero as the number of binding sites increases. This
constraint correlates with the general trend that the number of proteins
used to tie the DNA loop increases with the length of the loop \cite{ours,matthews}.
A longer loop typically implies a higher free energy of looping, $c$,
which in turn requires a stronger interaction between proteins (a
more negative $e$) or a higher number of sites in order for the system
to switch to the looped state.

A remarkable property inferred from the previous equations is that
the looping free energy and the number of binding sites affect the
concentration at which the transition occurs only through the ratio
$c/N$. If this ratio is kept constant, coordinated changes in $c$
and $N$ modify the sharpness of the transition but not the concentration
at which it happens (Figure \ref{cap:TO}). The main trends observed
in the looping behavior with respect to $c$ and $N$ are also observed
in the occupancy of the sites (Equation \ref{eq:Occupancy}), which
depends on $c$ and $N$ only through the looping probability.

In the case of large $N$, by expanding in terms of the dimensionless
parameter $\beta c/N$, the previous equation simplifies to\begin{equation}
\tilde{n}=e^{\beta g^{o}}\sqrt{\frac{c\beta}{N\left(e^{-e\beta}-1\right)}}\;,\label{eq:Sol-n-largeN}\end{equation}
 which indicates that the concentration at which the transition happens
decreases asymptotically like $N^{-1/2}$ as the number of binding
sites increases (as demonstrated in Figure \ref{cap:3D}a). Therefore,
by increasing the number of binding sites, the system can reach the
looped phase at arbitrarily small protein concentrations. 

This asymptotic equation indicates that, for large $N$, changes of
$e$ and $N$ that keep $N(e^{-e\beta}-1)$ constant do not affect
the transition point. Because of the strong dependence of the occupancy
on $e$ for looped phases (Equation \ref{eq:Occupancy}), coordinated
changes of $e$ and $N$ can keep the looping properties while strongly
affecting the occupancy of the sites (Figure \ref{cap:TS}).

In conclusion, we have developed the first model for DNA looping by
an arbitrary number of proteins and found that, for large number of
binding sites, the system exhibits a phase-transition-like behavior
with two phases in which DNA is either looped or unlooped. Many cellular
processes rely on the existence a looped phase to work (e.g. telomere
maintenance), others on the occupancies of the sites that comes with
the looped phase (e.g. gene regulation), and others on the transition
from one phase to another to trigger its effects (e.g. DNA recombination).
\emph{}Our results indicate that DNA looping by multiple proteins
has a high versatility to achieve different behaviors. Explicitly,
\emph{}the system can reach the looped phase at arbitrarily small
protein concentrations, the sharpness of the transition can easily
be tuned, and the system can choose the degree to which switching
to the looped state affects occupancy of the DNA binding sites. This
versatility underlies the many facets of DNA looping across the spectrum
of biological processes where it is at play.

The model we have proposed and its potential extensions encompass
a broad range of biological processes. The case of identical binding
we have discussed here in detail closely approximates DNA looping
in DNA recombination and telomere maintenance \cite{de Lange,broach}.
Both of these processes play a fundamental role in the functioning
of the cell and their deregulation is responsible for a variety of
diseases, including different types of cancer \cite{weinberg}. Our
model provides a backbone to build upon and to tackle more complex
situations, involving for instance non-identical binding, multiple
loops, and intra-operator interactions \cite{H general}. From a methodological
point of view, our approach provides a full Hamiltonian formulation
of DNA looping that opens the applicability of the techniques of statistical
physics, both computational and analytical, to a new range of biological
problems of basic and medical importance.


\begin{thebibliography}{10}
\bibitem{new ref}The typical values of the parameters $e$ and $c$ for DNA looping
in the natural environment of the cell fall in the range $e<0$ and
$c>0$ (attractive interactions between DNA-binding molecules and
DNA naturally unlooped). The model, analysis, and results are in general
valid for any value of the parameters $e$ and $c$. The case in which
$e>0$ and $c<0$ would mean that the loop is intrinsically stable
and the DNA-binding molecules {}``open'' the loop. For $e>0$ and
$c>0$, the system is always unlooped, and for $e<0$ and $c<0$,
the system is always looped.
\bibitem{H general}A more general Hamiltonian is given by: \begin{eqnarray*}
H & = & \sum_{k=1}^{M}\left[\sigma_{L,k}\left(c_{k}+\sum_{i,j=1}^{N}e_{i,j,k}\sigma_{U,i}\sigma_{D,j}\right)\right.\\
 &  & \left.+\sum_{i=1}^{N}\left(g_{U,i}\sigma_{U,i}+g_{D,i}\sigma_{D,i}\right)\right]\;,\end{eqnarray*}
where the binding free energy $g_{U/D,i}$ depends on the site, there
are $M$ different types of DNA loops with potentially different free
energies $c_{k}$, and the interaction free energy $e_{i,j,k}$ between
proteins bound at opposite operators depends of the type of loop and
the particular binding sites.
\end{thebibliography}
\end{document}